%% file: Aubry.tex
\documentclass[final]{svjour2}
\usepackage{graphicx}
\usepackage{rotating}
\usepackage{amssymb}
\usepackage{mathptmx}
\usepackage[numbers]{natbib}
\makeatletter
\journalname{Journal of Low Temperature Physics}

\bibpunct{}{}{,}{s}{}{,}

\begin{document}

\newcommand{\hdblarrow}{H\makebox[0.9ex][l]{$\downdownarrows$}-}
\title{Critical behavior of the liquid gas transition of $^4$He confined in a silica aerogel}

\author{G.J.~Aubry$^1$ \and F.~Bonnet$^1$ \and M.~Melich$^1$ \and L.~Guyon$^1$ \and F.~Despetis$^2$ \and P.~E.~Wolf$^1$}

\institute{1: Institut N\'eel CNRS-UJF, BP 166, F-38042 Grenoble
CEDEX 9, France\\
\email{pierre-etienne.wolf@grenoble.cnrs.fr}
\\2: Lab Charles Coulomb UMR 5221 CNRS-UM2,
F-34095 Montpellier CEDEX 5, France}

\date{11.07.2012}

\maketitle


\begin{abstract}
We have studied $^4$He confined in a 95\% porosity silica aerogel in
the vicinity of the bulk liquid gas critical point.  Both
thermodynamic measurements and light scattering experiments were
performed to probe the effect of a quenched disorder on the liquid gas
transition, in relation with the Random Field Ising Model (RFIM).  We
find that the hysteresis between condensation and evaporation present
at lower temperatures disappears at a temperature $T_{ch}$ between 25
and 30 mK below the critical point.  Slow relaxations are observed for
temperatures slightly below $T_{ch}$, indicating that some energy
barriers, but not all, can be overcome.  Above $T_{ch}$, no density
step is observed along the (reversible) isotherms, showing that the
critical behavior of the equilibrium phase transition in presence of
disorder, if it exists, is shifted to smaller temperatures, where it
cannot be observed due to the impossibility to reach equilibrium.
Above $T_{ch}$, light scattering exhibits a weak maximum close to the
pressure where the isotherm slope is maximal.  This behavior can be
accounted for by a simple model incorporating the compression of
$^4$He close to the silica strands.


%

\end{abstract}

\section{Introduction}

Despite intense theoretical and experimental efforts, the effect of
quenched disorder on phase transitions remains not fully understood.
While this question has originally arisen in the context of magnetism,
it has been extended to fluids confined in disordered porous materials,
in particular tenuous silica gels or aerogels which offer an unique
experimental realization of a self-sustained, dilute, quenched
disorder.  Focusing on the Ising universality class, Brochard and de
Gennes have suggested that the demixtion of a binary liquid confined
in a gel should be an experimental realization of the Random Field
Ising Model (RFIM) and of its critical behavior
\cite{DeGennesBrochard83}, at least close enough to the bulk critical
temperature $T_{c}$ for the fluid correlation length to be larger than
the gel correlation length.  In the same spirit, a number of studies
have concerned the liquid-gas transition of $^4$He confined in
aerogels.  Early experiments by Wong and Chan\cite{WongChanPRL90} have
been interpreted in terms of a genuine first-order equilibrium phase
transition existing in a $\it{wide}$ temperature range, with a shifted
critical point and a modified critical exponent for the order
parameter (the density difference between liquid and gas), due to the
disorder.  However, latter experiments on $^4$He in aerogels
\cite{GabayPhysica00,Lambert:2004zr,HermanPRB05,BonnetEPL08} have
revealed a strong hysteresis between condensation and evaporation,
persisting up to close to the critical temperature \cite{HermanPRB05}
, indicating that the system cannot reach equilibrium.  A remarkable
feature is that, at low temperatures or large porosities, the
adsorption isotherm presents a quasi-plateau of pressure, just like
for a regular first-order, equilibrium, transition
\cite{Lambert:2004zr,HermanPRB05,BonnetEPL08, Tulimieri99}.  In fact,
this behavior turns out to be consistent \cite{BonnetEPL08}
with the prediction originally made for the zero temperature RFIM
\cite{Sethna93}, and generalized to the case of the liquid-gas
transition at finite temperature \cite{Kierlik01,
Detcheverry03, Detcheverry04}, of a disorder-driven, non equilibrium,
phase transition, where condensation occurs by athermal avalanches,
whose size diverges below a critical temperature (or disorder).

These results leave open the possibility that, very close to $T_{c}$,
the energy barriers for condensation become small enough for the
system to be described by the equilibrium RFIM, characterized by a
true phase transition below a critical temperature $T_{c}^{*}$, and
non bulk critical exponents at $T_{c}^{*}$ \cite{YoungJPA1993}.
Experiments on N$_2$ \cite{WongPRL93} and isobutyric acid-water
\cite{ZhuCannelPRL96} confined in (aero)gels have been interpreted as
showing such an equilibrium behavior, while the observation of slow
dynamics above $T_{c}^{*}$ in these experiments fitted with the
initial theoretical suggestion that, above $T_{c}^{*}$, the Ising RFIM
model could present an equilibrium glassy phase \cite{MezardPRB1994}.
However, it has been recently rigorously shown that only an out of
equilibrium glassy phase is possible \cite{Zdeberova2010} above
$T_{c}^{*}$, and it is not clear how the experiments fit this picture.

In contrast, two other experiments find no evidence for any
equilibrium critical behavior.  For $^4$He confined in aerogels,
isotherms measurements very close to $T_{c}$, in the region where the
hysteresis disappears, show no true pressure plateau, \textit{i.e.} no
indication of an equilibrium phase transition \cite{HermanPRB05}.  On the other hand,
neutron scattering measurements performed on CO$_2$ confined in a
silica aerogel close to the bulk critical
point \cite{MelnichenkoPRE2004,MelnichenkoJCP2006} show that the fluid
correlation length remains finite, consistent with a suppression of
the phase transition due to disorder.  In this paper, we revisit this
question by combining, in a single experiment, high resolution
isotherms \emph{and} light scattering measurements for $^4$He confined
in aerogels close to $T_{c}$.
\section{Experimental}
We have studied two aerogels samples with the same porosity (95\%) and
different microscopic structures, obtained respectively by synthesis
in basic (B102) or neutral (N102) conditions.  Both samples are about
3.7 mm thick and 14~mm in diameter and located in an optical cell,
allowing light scattering measurements at 45$^\circ$ and 135$^\circ$
from the incident direction \cite{BonnetEPL08}.  The cell temperature
is measured by a germanium resistor, and regulated to about 10 $\mu$K
by a heater glued onto the cell walls.  Condensation is controlled by
changing between 80 K and 200 K the temperature of an external
reservoir connected to the cell \cite{Cross2007}.  The mass condensed
into the aerogel is precisely obtained from the total amount of $^4$He
in the system and the reservoir temperature \cite{BonnetEPL08}.  The
pressure is measured by a room temperature sensor connected to the
cell by a separate capillary.  During condensation, latent heat is
released and flows to the cell walls, causing a local increase of the
$^4$He temperature, due to its poor thermal conductivity.  In order to
suppress this thermal gradient, we stop condensation at regular time
intervals and let the system thermally relax about one hour while
continuously measuring the pressure $P$ and the heating power $W_{h}$
necessary to regulate the cell temperature.

\section{Isotherms}\label{isotherms}

\begin{figure}
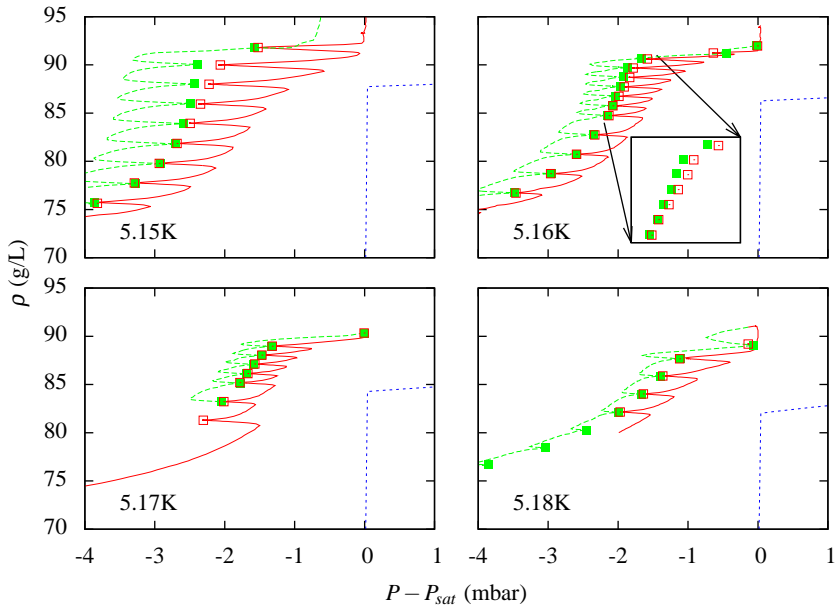

  \centering
  \include{120705_isothermes}

  \caption{(Color online) Isotherms in B102 close to $T_c$.  The solid
  line (red online) is the adsorption branch, while the dashed line
  (green online) is the desorption branch.  The dashed line in the
  lower right corner (blue online) represents the bulk isotherm
  \cite{Kierstead73} for the given temperature. Bulk gas densities are out of range on the scale of this figure and fall between 52~g/L (5.15~K) and 57~g/L (5.18~K). Hysteresis closes
  between 5.16~K and 5.17~K.}
  \label{fig:isothermes}
\end{figure}

Figure~\ref{fig:isothermes} shows isotherms for B102 close to
$T_{c}$=5.195 K, for condensation or evaporation rates between 0.25 and
0.50~$\mu$mol/s.  Each stop is followed by a relaxation of the
pressure.  Once the pressure has relaxed, the loop remains open at
5.15~K and 5.16~K while it is closed at 5.17 K and 5.18 K within our
pressure resolution (10 $\mu$bar).  Hence, the hysteresis loop closes
between 25 and 35~mK below $T_{c}$, in agreement with the results of
Herman \textit{et al.} on a similar aerogel \cite{HermanPRB05}.  This
is intermediate between the temperatures (respectively 9 and 65~mK
below $T_{c}$) where the bulk correlation length $\xi$
\cite{RoeJLTP78}, and the interface width (3.6 $\xi$ \cite{bonnJP92})
become equal to the estimated aerogel correlation length of 10~nm
\cite{BonnetEPL08} .  For the two temperatures between the closure
temperature $T_{ch}$ and $T_{c}$, the isotherm does not present any
pressure plateau.  As Herman \textit{et al.}\cite{HermanPRB05}, we do
dot find any evidence for any equilibrium long range order below
$T_{c}$.

The isotherms present a kink at a certain density, above which the
pressure rises faster, with a slope comparable to that of the bulk
isotherms just above the saturated pressure, showing that the kink
marks the end of the condensation inside the aerogel.  The
corresponding density of the confined liquid is only slightly larger
than the bulk density (at most 10\%).  While this agrees with
ref.\cite{HermanPRB05} and the fact that most of $^4$He is far
enough from the silica strands not to be compressed by the Van der
Waals interaction, this contrasts with the conclusion of
ref.\cite{MelnichenkoJCP2006}, based on neutrons transmission measurements,
that the average density of CO$_2$ in aerogel close to the critical point
is 50\% larger than in the bulk. We will come back to this point in
\S\ref{optics}.

\section{Pressure relaxation and slow dynamics}
\begin{figure}
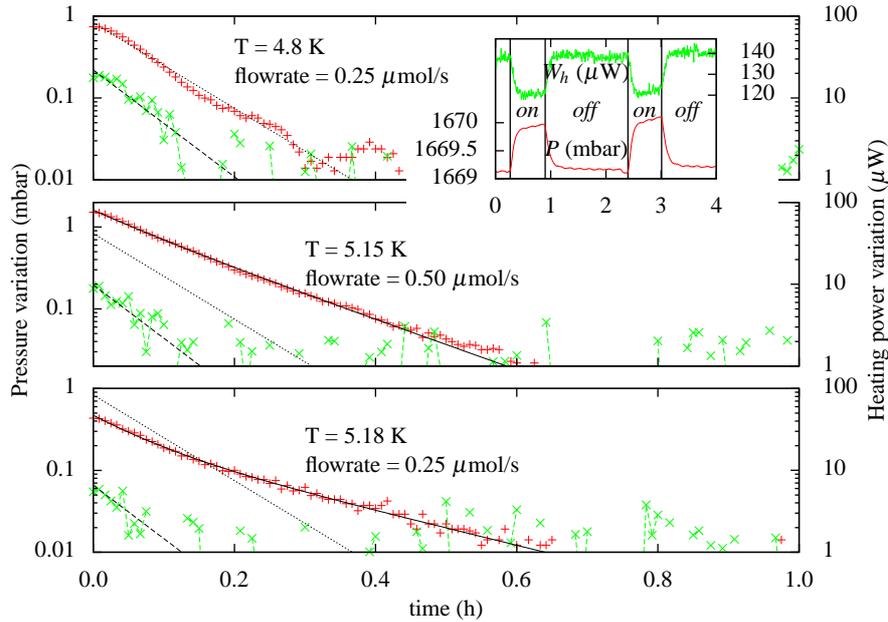

  \centering
  \include{120706_4p8-5p15-5p18}
  \caption{(Color online)
 Relaxations of the pressure $P$ ($+$) and of the heater power $W_{h}$
 ($\times$) following a condensation step at three temperatures.  The
 inset shows $P$ and $W_{h}$ at 4.8~K with the flow rate \textit{on} or \textit{off}.
 The behavior at 4.8~K is fitted with single exponentials and is
 attributed to thermal relaxation.  For 5.15~K and 5.18~K, the
 dashed and dotted lines are exponentials with the decay rates
 measured at 4.8~K for $P$ and $W_{h}$.  Pressure relaxation of $W_{h}$
 at 5.15~K and 5.18~K is slower than at 4.8~K and can be fitted
 (solid lines) by the pressure relaxation at 4.8~K, rescaled by the
 amplitude of $W_{h}$, plus a slower exponential, which we attribute
 to slow dynamics of the fluid distribution near $T_{ch}$.}
  \label{fig:relaxations}
\end{figure}

We have searched for slow dynamics characteristic of disordered
systems by examining in detail the relaxation of pressure following
each condensation step. At 4.8~K, far below $T_c$, figure~\ref{fig:relaxations}(a) shows that once the condensation is
stopped, $P$ and $W_h$ relax to their final values with a similar time scale of about 5 minutes. This is comparable to the
computed decay time (2 minutes) of the slowest mode of thermal
diffusion in a 3.7 mm thick slab filled with bulk liquid, suggesting
that the observed pressure relaxation could be related to the cooling
of $^4$He following the stop of condensation.  Indeed, we have
checked that decreasing the cell temperature at constant
filling on the pressure quasi-plateau does shift the measured
pressure by the corresponding change in the saturated vapor pressure
(about 1.3 mbar per mK).

Figures~\ref{fig:relaxations}(b) and (c) show relaxations close to
$T_{c}$, below and above $T_{ch}$.  $W_{h}$ relaxes on the same time
scale of 5 minutes with an amplitude proportional to the initial flow
rate, consistent with a thermal effect similar to that at 4.8~K. In
contrast, the pressure relaxes on a longer time scale, which depends
on the average density and is maximal close to full condensation.  We
can fit the observed relaxation by a sum of two exponentials, one
proportional to $W_{h}$ due to thermal relaxation, the other with a
slower decay rate (about 8 minutes at 5.15~K and 12 minutes at 5.18~K
close to full condensation) which could reflect slow dynamics of the
fluid distribution inside the aerogel.  This would be consistent with
the existence of energy barriers.  The longer relaxation time at
5.18~K could result from the fact that unsurmountable barriers below
$T_{ch}$ can be overcome on the observation time scale.
However, on the one hour timescale of our measurements, we do not
observe the stretched exponential behavior expected for a glassy
dynamics associated with a distribution of barriers.  Direct
measurements of the fluid dynamics using photon correlation
spectroscopy would be needed to elucidate whether this feature could
be due to our limited pressure resolution.
\section{Light scattering close to $T_{c}$ }\label{optics}
\begin{figure}
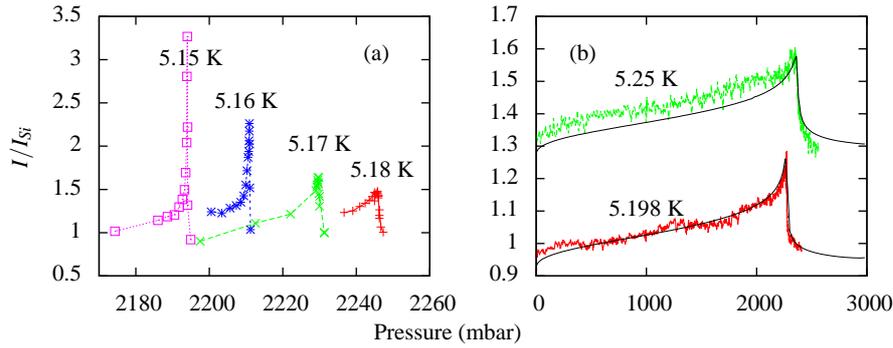

  \centering
  \include{120710_diffusion}
  \caption{(Color online) Pressure
  dependence of the light intensity scattered at 45$^\circ$ for B102
  (a) and 135$^\circ$ for N102 (b), normalized by the intensity
  $I_{Si}$ scattered by the bare aerogel.  In figure (b), the curve at 5.25~K is  vertically shifted by 0.3 for clarity.  Figure (b) also shows the
  theoretical dependence of the excess mass adsorbed on a flat silica
  substrate, rescaled for comparison to the optical signal (see text).}
  \label{fig:scattering}
\end{figure}
Figure~\ref{fig:scattering}(a) shows the intensity scattered at
45$^\circ$ as a function of pressure for B102.
For $T \leq T_{ch}$, this signal (as well as that at 135$^\circ$)
increases in the region of the hysteresis loop.  In N102, the light
scattered by the entrance region of the laser sheet at 135$^\circ$
shows a similar behavior (the light scattered by other regions, or at
45$^\circ$, is dominated by the strong silica background of N102
\cite{BonnetEPL08}).  As at lower temperatures \cite{BonnetEPL08},
this behavior is consistent with the progressive growth of
microscopic static liquid domains as the pressure increases.  The
decrease of the signal with temperature is consistent with the
decrease in the optical contrast between liquid and gas as the bulk
critical temperature is approached.  In this interpretation, one could
expect the optical signal to vanish at $T_{c}$.  However, measurements
in N102 performed 3 and 50~mK above $T_{c}$ show a continuous increase
of the scattered intensity up to the region of maximal slope in the
bulk pressure-density isotherm, followed by an abrupt drop to the
background silica level (Figure~\ref{fig:scattering}(b)).
Correlatively, the measured transmission drops in the region of the peak in the
scattered intensity (the logarithm of the extra attenuation with
respect to the bare silica being proportional to the scattered signal
at 135$^\circ$, as expected).  While this behavior could evocate
critical opalescence, it can be accounted for by the concept of critical
adsorption \cite{FisherDeGennesCRAS1978}.  The van der Waals
attraction by the silica, combined with the large bulk compressibility
of $^4$He close to $T_{c}$, increases the density of $^4$He close to
the silica strands, hence the scattered signal, in a similar way to
the case of an adsorbed film \cite{Wolf2009epje}.  With respect to an
homogeneous situation, the scattered field is increased by a quantity
proportional to the excess mass with respect to the bulk, defined as
the volume averaged difference between the local density and the bulk
density at the external pressure.  This quantity can be computed as a function of temperature in
the case of a planar substrate, using the bulk state equation of
$^4$He \cite{Kierstead73} and the van der Waals attraction given in
ref.\cite{SaamColePRB75}.  As shown by figure~\ref{fig:scattering}(b),
the temperature dependence of the obtained excess mass reproduces
quite well the observed scattered intensity, the scaling factor being
close to that computed from the known
dielectric constants of silica and $^4$He, and a typical diameter strand of 2~nm.  On the other hand, the
measured extra scattering due to $^4$He (about 25\% of the silica
contribution, corresponding to a mean free path of 10~mm) is nearly
thousand times larger than the (separately measured) bulk scattering
signal at $T_{c}$ and 45$^\circ$.  Here also, photon correlation
spectroscopy would be needed to separate possible critical thermodynamic
temporal fluctuations from the strong static contribution of critical
adsorption.

Finally, we note that our results for N102 are quite similar to those
obtained on CO$_2$ using neutrons rather than light scattering\cite{MelnichenkoJCP2006}.  This
suggests that, in the latter case, the decrease in transmission
measured near $T_{c}$ close to the critical pressure is also due to
the increase of scattering due to critical adsorption.  Extracting the
fluid density from the neutrons transmission, as the authors of
ref.\cite{MelnichenkoJCP2006} do, would then be incorrect.  This might be the
origin of the reported large increase of the confined density with
respect to the bulk fluid for CO$_2$ (see \S\ref{isotherms}).

\section{Conclusions}
In summary, in contrast to
refs.\cite{WongChanPRL90,WongPRL93,ZhuCannelPRL96}, our experimental
results do not show any equilibrium first order transition above
$T_{ch}$, the closure temperature of the hysteresis loop.  Either, the
transition predicted by the RFIM lies inside the hysteretic region
below $T_{ch}$, or the temperature above which the bulk $^4$He
correlation length is large enough for the RFIM to apply is larger
than $T_{c}^{*}$, the critical temperature of the RFIM model.  On the
other hand, around $T_{ch}$, the relaxation of pressure does not show
any clear clear evidence for marked glassy dynamics.  Finally, close
and above the bulk critical temperature, the optical signal is
dominated by the static contribution resulting from the large
compressibility of $^4$He, masking any possible critical dynamic
contribution.  To go further, we plan to improve the sensitivity of
our measurements by using photon correlation spectroscopy to directly
detect small dynamic changes of the fluid distribution.

%

\end{document}

%% file: 120705_isothermes.tex
\begingroup
  \makeatletter
  \providecommand\color[2][]{%
    \GenericError{(gnuplot) \space\space\space\@spaces}{%
      Package color not loaded in conjunction with
      terminal option `colourtext'%
    }{See the gnuplot documentation for explanation.%
    }{Either use 'blacktext' in gnuplot or load the package
      color.sty in LaTeX.}%
    \renewcommand\color[2][]{}%
  }%
  \providecommand\includegraphics[2][]{%
    \GenericError{(gnuplot) \space\space\space\@spaces}{%
      Package graphicx or graphics not loaded%
    }{See the gnuplot documentation for explanation.%
    }{The gnuplot epslatex terminal needs graphicx.sty or graphics.sty.}%
    \renewcommand\includegraphics[2][]{}%
  }%
  \providecommand\rotatebox[2]{#2}%
  \@ifundefined{ifGPcolor}{%
    \newif\ifGPcolor
    \GPcolortrue
  }{}%
  \@ifundefined{ifGPblacktext}{%
    \newif\ifGPblacktext
    \GPblacktexttrue
  }{}%
  \let\gplgaddtomacro\g@addto@macro
  \gdef\gplbacktext{}%
  \gdef\gplfronttext{}%
  \makeatother
  \ifGPblacktext
    \def\colorrgb#1{}%
    \def\colorgray#1{}%
  \else
    \ifGPcolor
      \def\colorrgb#1{\color[rgb]{#1}}%
      \def\colorgray#1{\color[gray]{#1}}%
      \expandafter\def\csname LTw\endcsname{\color{white}}%
      \expandafter\def\csname LTb\endcsname{\color{black}}%
      \expandafter\def\csname LTa\endcsname{\color{black}}%
      \expandafter\def\csname LT0\endcsname{\color[rgb]{1,0,0}}%
      \expandafter\def\csname LT1\endcsname{\color[rgb]{0,1,0}}%
      \expandafter\def\csname LT2\endcsname{\color[rgb]{0,0,1}}%
      \expandafter\def\csname LT3\endcsname{\color[rgb]{1,0,1}}%
      \expandafter\def\csname LT4\endcsname{\color[rgb]{0,1,1}}%
      \expandafter\def\csname LT5\endcsname{\color[rgb]{1,1,0}}%
      \expandafter\def\csname LT6\endcsname{\color[rgb]{0,0,0}}%
      \expandafter\def\csname LT7\endcsname{\color[rgb]{1,0.3,0}}%
      \expandafter\def\csname LT8\endcsname{\color[rgb]{0.5,0.5,0.5}}%
    \else
      \def\colorrgb#1{\color{black}}%
      \def\colorgray#1{\color[gray]{#1}}%
      \expandafter\def\csname LTw\endcsname{\color{white}}%
      \expandafter\def\csname LTb\endcsname{\color{black}}%
      \expandafter\def\csname LTa\endcsname{\color{black}}%
      \expandafter\def\csname LT0\endcsname{\color{black}}%
      \expandafter\def\csname LT1\endcsname{\color{black}}%
      \expandafter\def\csname LT2\endcsname{\color{black}}%
      \expandafter\def\csname LT3\endcsname{\color{black}}%
      \expandafter\def\csname LT4\endcsname{\color{black}}%
      \expandafter\def\csname LT5\endcsname{\color{black}}%
      \expandafter\def\csname LT6\endcsname{\color{black}}%
      \expandafter\def\csname LT7\endcsname{\color{black}}%
      \expandafter\def\csname LT8\endcsname{\color{black}}%
    \fi
  \fi
  \setlength{\unitlength}{0.0500bp}%
  \begin{picture}(6518.00,4534.00)%
    \gplgaddtomacro\gplbacktext{%
      \csname LTb\endcsname%
      \put(519,2607){\makebox(0,0)[r]{\strut{} 70}}%
      \put(519,2969){\makebox(0,0)[r]{\strut{} 75}}%
      \put(519,3332){\makebox(0,0)[r]{\strut{} 80}}%
      \put(519,3694){\makebox(0,0)[r]{\strut{} 85}}%
      \put(519,4057){\makebox(0,0)[r]{\strut{} 90}}%
      \put(519,4419){\makebox(0,0)[r]{\strut{} 95}}%
      \put(651,2387){\makebox(0,0){\strut{}}}%
      \put(1172,2387){\makebox(0,0){\strut{}}}%
      \put(1694,2387){\makebox(0,0){\strut{}}}%
      \put(2215,2387){\makebox(0,0){\strut{}}}%
      \put(2737,2387){\makebox(0,0){\strut{}}}%
      \put(3258,2387){\makebox(0,0){\strut{}}}%
      \put(912,2788){\makebox(0,0)[l]{\strut{}5.15K}}%
      \put(163,2493){\rotatebox{-270}{\makebox(0,0){\strut{}$\rho$ (g/L)}}}%
      \put(3421,68){\makebox(0,0){\strut{}$P-P_{sat}$ (mbar)}}%
    }%
    \gplgaddtomacro\gplfronttext{%
    }%
    \gplgaddtomacro\gplbacktext{%
      \csname LTb\endcsname%
      \put(3452,2607){\makebox(0,0)[r]{\strut{}}}%
      \put(3452,2969){\makebox(0,0)[r]{\strut{}}}%
      \put(3452,3332){\makebox(0,0)[r]{\strut{}}}%
      \put(3452,3694){\makebox(0,0)[r]{\strut{}}}%
      \put(3452,4057){\makebox(0,0)[r]{\strut{}}}%
      \put(3452,4419){\makebox(0,0)[r]{\strut{}}}%
      \put(3584,2387){\makebox(0,0){\strut{}}}%
      \put(4105,2387){\makebox(0,0){\strut{}}}%
      \put(4627,2387){\makebox(0,0){\strut{}}}%
      \put(5148,2387){\makebox(0,0){\strut{}}}%
      \put(5670,2387){\makebox(0,0){\strut{}}}%
      \put(6191,2387){\makebox(0,0){\strut{}}}%
      \put(3845,2788){\makebox(0,0)[l]{\strut{}5.16K}}%
    }%
    \gplgaddtomacro\gplfronttext{%
    }%
    \gplgaddtomacro\gplbacktext{%
    }%
    \gplgaddtomacro\gplfronttext{%
    }%
    \gplgaddtomacro\gplbacktext{%
      \csname LTb\endcsname%
      \put(519,566){\makebox(0,0)[r]{\strut{} 70}}%
      \put(519,929){\makebox(0,0)[r]{\strut{} 75}}%
      \put(519,1291){\makebox(0,0)[r]{\strut{} 80}}%
      \put(519,1654){\makebox(0,0)[r]{\strut{} 85}}%
      \put(519,2016){\makebox(0,0)[r]{\strut{} 90}}%
      \put(519,2379){\makebox(0,0)[r]{\strut{} 95}}%
      \put(651,346){\makebox(0,0){\strut{}-4}}%
      \put(1172,346){\makebox(0,0){\strut{}-3}}%
      \put(1694,346){\makebox(0,0){\strut{}-2}}%
      \put(2215,346){\makebox(0,0){\strut{}-1}}%
      \put(2737,346){\makebox(0,0){\strut{} 0}}%
      \put(3258,346){\makebox(0,0){\strut{} 1}}%
      \put(912,747){\makebox(0,0)[l]{\strut{}5.17K}}%
    }%
    \gplgaddtomacro\gplfronttext{%
    }%
    \gplgaddtomacro\gplbacktext{%
      \csname LTb\endcsname%
      \put(3452,566){\makebox(0,0)[r]{\strut{}}}%
      \put(3452,929){\makebox(0,0)[r]{\strut{}}}%
      \put(3452,1291){\makebox(0,0)[r]{\strut{}}}%
      \put(3452,1654){\makebox(0,0)[r]{\strut{}}}%
      \put(3452,2016){\makebox(0,0)[r]{\strut{}}}%
      \put(3452,2379){\makebox(0,0)[r]{\strut{}}}%
      \put(3584,346){\makebox(0,0){\strut{}-4}}%
      \put(4105,346){\makebox(0,0){\strut{}-3}}%
      \put(4627,346){\makebox(0,0){\strut{}-2}}%
      \put(5148,346){\makebox(0,0){\strut{}-1}}%
      \put(5670,346){\makebox(0,0){\strut{} 0}}%
      \put(6191,346){\makebox(0,0){\strut{} 1}}%
      \put(3845,747){\makebox(0,0)[l]{\strut{}5.18K}}%
    }%
    \gplgaddtomacro\gplfronttext{%
    }%
    \gplbacktext
    \put(0,0){\includegraphics{120705_isothermes.eps_tex}}%
    \gplfronttext
  \end{picture}%
\endgroup

%% file: 120706_4p8-5p15-5p18.tex
\begingroup
  \makeatletter
  \providecommand\color[2][]{%
    \GenericError{(gnuplot) \space\space\space\@spaces}{%
      Package color not loaded in conjunction with
      terminal option `colourtext'%
    }{See the gnuplot documentation for explanation.%
    }{Either use 'blacktext' in gnuplot or load the package
      color.sty in LaTeX.}%
    \renewcommand\color[2][]{}%
  }%
  \providecommand\includegraphics[2][]{%
    \GenericError{(gnuplot) \space\space\space\@spaces}{%
      Package graphicx or graphics not loaded%
    }{See the gnuplot documentation for explanation.%
    }{The gnuplot epslatex terminal needs graphicx.sty or graphics.sty.}%
    \renewcommand\includegraphics[2][]{}%
  }%
  \providecommand\rotatebox[2]{#2}%
  \@ifundefined{ifGPcolor}{%
    \newif\ifGPcolor
    \GPcolortrue
  }{}%
  \@ifundefined{ifGPblacktext}{%
    \newif\ifGPblacktext
    \GPblacktexttrue
  }{}%
  \let\gplgaddtomacro\g@addto@macro
  \gdef\gplbacktext{}%
  \gdef\gplfronttext{}%
  \makeatother
  \ifGPblacktext
    \def\colorrgb#1{}%
    \def\colorgray#1{}%
  \else
    \ifGPcolor
      \def\colorrgb#1{\color[rgb]{#1}}%
      \def\colorgray#1{\color[gray]{#1}}%
      \expandafter\def\csname LTw\endcsname{\color{white}}%
      \expandafter\def\csname LTb\endcsname{\color{black}}%
      \expandafter\def\csname LTa\endcsname{\color{black}}%
      \expandafter\def\csname LT0\endcsname{\color[rgb]{1,0,0}}%
      \expandafter\def\csname LT1\endcsname{\color[rgb]{0,1,0}}%
      \expandafter\def\csname LT2\endcsname{\color[rgb]{0,0,1}}%
      \expandafter\def\csname LT3\endcsname{\color[rgb]{1,0,1}}%
      \expandafter\def\csname LT4\endcsname{\color[rgb]{0,1,1}}%
      \expandafter\def\csname LT5\endcsname{\color[rgb]{1,1,0}}%
      \expandafter\def\csname LT6\endcsname{\color[rgb]{0,0,0}}%
      \expandafter\def\csname LT7\endcsname{\color[rgb]{1,0.3,0}}%
      \expandafter\def\csname LT8\endcsname{\color[rgb]{0.5,0.5,0.5}}%
    \else
      \def\colorrgb#1{\color{black}}%
      \def\colorgray#1{\color[gray]{#1}}%
      \expandafter\def\csname LTw\endcsname{\color{white}}%
      \expandafter\def\csname LTb\endcsname{\color{black}}%
      \expandafter\def\csname LTa\endcsname{\color{black}}%
      \expandafter\def\csname LT0\endcsname{\color{black}}%
      \expandafter\def\csname LT1\endcsname{\color{black}}%
      \expandafter\def\csname LT2\endcsname{\color{black}}%
      \expandafter\def\csname LT3\endcsname{\color{black}}%
      \expandafter\def\csname LT4\endcsname{\color{black}}%
      \expandafter\def\csname LT5\endcsname{\color{black}}%
      \expandafter\def\csname LT6\endcsname{\color{black}}%
      \expandafter\def\csname LT7\endcsname{\color{black}}%
      \expandafter\def\csname LT8\endcsname{\color{black}}%
    \fi
  \fi
  \setlength{\unitlength}{0.0500bp}%
  \begin{picture}(6518.00,4534.00)%
    \gplgaddtomacro\gplbacktext{%
      \csname LTb\endcsname%
      \put(3225,2733){\makebox(0,0){\strut{}}}%
      \put(65,2267){\rotatebox{-270}{\makebox(0,0){\strut{}Pressure variation (mbar)}}}%
      \put(6452,2267){\rotatebox{-270}{\makebox(0,0){\strut{}Heating power variation ($\mu$W)}}}%
    }%
    \gplgaddtomacro\gplfronttext{%
      \csname LTb\endcsname%
      \put(462,3019){\makebox(0,0)[r]{\strut{} 0.01}}%
      \put(462,3635){\makebox(0,0)[r]{\strut{} 0.1}}%
      \put(462,4250){\makebox(0,0)[r]{\strut{} 1}}%
      \put(594,2799){\makebox(0,0){\strut{}}}%
      \put(1647,2799){\makebox(0,0){\strut{}}}%
      \put(2699,2799){\makebox(0,0){\strut{}}}%
      \put(3752,2799){\makebox(0,0){\strut{}}}%
      \put(4804,2799){\makebox(0,0){\strut{}}}%
      \put(5857,2799){\makebox(0,0){\strut{}}}%
      \put(5989,3019){\makebox(0,0)[l]{\strut{} 1}}%
      \put(5989,3635){\makebox(0,0)[l]{\strut{} 10}}%
      \put(5989,4250){\makebox(0,0)[l]{\strut{} 100}}%
      \put(1647,4004){\makebox(0,0)[l]{\strut{}T = 4.8 K}}%
      \put(1647,3784){\makebox(0,0)[l]{\strut{}flowrate = 0.25 $\mu$mol/s}}%
    }%
    \gplgaddtomacro\gplbacktext{%
      \put(3225,1333){\makebox(0,0){\strut{}}}%
    }%
    \gplgaddtomacro\gplfronttext{%
      \csname LTb\endcsname%
      \put(462,2049){\makebox(0,0)[r]{\strut{} 0.1}}%
      \put(462,2665){\makebox(0,0)[r]{\strut{} 1}}%
      \put(594,1399){\makebox(0,0){\strut{}}}%
      \put(1647,1399){\makebox(0,0){\strut{}}}%
      \put(2699,1399){\makebox(0,0){\strut{}}}%
      \put(3752,1399){\makebox(0,0){\strut{}}}%
      \put(4804,1399){\makebox(0,0){\strut{}}}%
      \put(5857,1399){\makebox(0,0){\strut{}}}%
      \put(5989,1619){\makebox(0,0)[l]{\strut{} 1}}%
      \put(5989,2235){\makebox(0,0)[l]{\strut{} 10}}%
      \put(5989,2850){\makebox(0,0)[l]{\strut{} 100}}%
      \put(2173,2481){\makebox(0,0)[l]{\strut{}T = 5.15 K}}%
      \put(2173,2261){\makebox(0,0)[l]{\strut{}flowrate = 0.50 $\mu$mol/s}}%
    }%
    \gplgaddtomacro\gplbacktext{%
      \put(3225,-220){\makebox(0,0){\strut{}time (h)}}%
    }%
    \gplgaddtomacro\gplfronttext{%
      \csname LTb\endcsname%
      \put(462,220){\makebox(0,0)[r]{\strut{} 0.01}}%
      \put(462,835){\makebox(0,0)[r]{\strut{} 0.1}}%
      \put(462,1451){\makebox(0,0)[r]{\strut{} 1}}%
      \put(594,0){\makebox(0,0){\strut{}0.0}}%
      \put(1647,0){\makebox(0,0){\strut{}0.2}}%
      \put(2699,0){\makebox(0,0){\strut{}0.4}}%
      \put(3752,0){\makebox(0,0){\strut{}0.6}}%
      \put(4804,0){\makebox(0,0){\strut{}0.8}}%
      \put(5857,0){\makebox(0,0){\strut{}1.0}}%
      \put(5989,220){\makebox(0,0)[l]{\strut{} 1}}%
      \put(5989,836){\makebox(0,0)[l]{\strut{} 10}}%
      \put(5989,1451){\makebox(0,0)[l]{\strut{} 100}}%
      \put(2173,1082){\makebox(0,0)[l]{\strut{}T = 5.18 K}}%
      \put(2173,862){\makebox(0,0)[l]{\strut{}flowrate = 0.25 $\mu$mol/s}}%
    }%
    \gplgaddtomacro\gplbacktext{%
    }%
    \gplgaddtomacro\gplfronttext{%
      \csname LTb\endcsname%
      \put(3461,3028){\makebox(0,0)[r]{\strut{} 1669}}%
      \put(3461,3238){\makebox(0,0)[r]{\strut{} 1669.5}}%
      \put(3461,3449){\makebox(0,0)[r]{\strut{} 1670}}%
      \put(3593,2808){\makebox(0,0){\strut{}0}}%
      \put(4005,2808){\makebox(0,0){\strut{}1}}%
      \put(4417,2808){\makebox(0,0){\strut{}2}}%
      \put(4828,2808){\makebox(0,0){\strut{}3}}%
      \put(5240,2808){\makebox(0,0){\strut{}4}}%
      \put(5372,3656){\makebox(0,0)[l]{\strut{} 120}}%
      \put(5372,3813){\makebox(0,0)[l]{\strut{} 130}}%
      \put(5372,3970){\makebox(0,0)[l]{\strut{} 140}}%
      \put(3833,3528){\makebox(0,0){\strut{}\textit{on}}}%
      \put(4272,3528){\makebox(0,0){\strut{}\textit{off}}}%
      \put(4705,3528){\makebox(0,0){\strut{}\textit{on}}}%
      \put(5034,3528){\makebox(0,0){\strut{}\textit{off}}}%
      \put(4272,3813){\makebox(0,0){\strut{}$W_h$ ($\mu$W)}}%
      \put(4272,3238){\makebox(0,0){\strut{}$P$ (mbar)}}%
    }%
    \gplbacktext
    \put(0,0){\includegraphics{120706_4p8-5p15-5p18.eps_tex}}%
    \gplfronttext
  \end{picture}%
\endgroup

%% file: 120710_diffusion.tex
\begingroup
  \makeatletter
  \providecommand\color[2][]{%
    \GenericError{(gnuplot) \space\space\space\@spaces}{%
      Package color not loaded in conjunction with
      terminal option `colourtext'%
    }{See the gnuplot documentation for explanation.%
    }{Either use 'blacktext' in gnuplot or load the package
      color.sty in LaTeX.}%
    \renewcommand\color[2][]{}%
  }%
  \providecommand\includegraphics[2][]{%
    \GenericError{(gnuplot) \space\space\space\@spaces}{%
      Package graphicx or graphics not loaded%
    }{See the gnuplot documentation for explanation.%
    }{The gnuplot epslatex terminal needs graphicx.sty or graphics.sty.}%
    \renewcommand\includegraphics[2][]{}%
  }%
  \providecommand\rotatebox[2]{#2}%
  \@ifundefined{ifGPcolor}{%
    \newif\ifGPcolor
    \GPcolortrue
  }{}%
  \@ifundefined{ifGPblacktext}{%
    \newif\ifGPblacktext
    \GPblacktexttrue
  }{}%
  \let\gplgaddtomacro\g@addto@macro
  \gdef\gplbacktext{}%
  \gdef\gplfronttext{}%
  \makeatother
  \ifGPblacktext
    \def\colorrgb#1{}%
    \def\colorgray#1{}%
  \else
    \ifGPcolor
      \def\colorrgb#1{\color[rgb]{#1}}%
      \def\colorgray#1{\color[gray]{#1}}%
      \expandafter\def\csname LTw\endcsname{\color{white}}%
      \expandafter\def\csname LTb\endcsname{\color{black}}%
      \expandafter\def\csname LTa\endcsname{\color{black}}%
      \expandafter\def\csname LT0\endcsname{\color[rgb]{1,0,0}}%
      \expandafter\def\csname LT1\endcsname{\color[rgb]{0,1,0}}%
      \expandafter\def\csname LT2\endcsname{\color[rgb]{0,0,1}}%
      \expandafter\def\csname LT3\endcsname{\color[rgb]{1,0,1}}%
      \expandafter\def\csname LT4\endcsname{\color[rgb]{0,1,1}}%
      \expandafter\def\csname LT5\endcsname{\color[rgb]{1,1,0}}%
      \expandafter\def\csname LT6\endcsname{\color[rgb]{0,0,0}}%
      \expandafter\def\csname LT7\endcsname{\color[rgb]{1,0.3,0}}%
      \expandafter\def\csname LT8\endcsname{\color[rgb]{0.5,0.5,0.5}}%
    \else
      \def\colorrgb#1{\color{black}}%
      \def\colorgray#1{\color[gray]{#1}}%
      \expandafter\def\csname LTw\endcsname{\color{white}}%
      \expandafter\def\csname LTb\endcsname{\color{black}}%
      \expandafter\def\csname LTa\endcsname{\color{black}}%
      \expandafter\def\csname LT0\endcsname{\color{black}}%
      \expandafter\def\csname LT1\endcsname{\color{black}}%
      \expandafter\def\csname LT2\endcsname{\color{black}}%
      \expandafter\def\csname LT3\endcsname{\color{black}}%
      \expandafter\def\csname LT4\endcsname{\color{black}}%
      \expandafter\def\csname LT5\endcsname{\color{black}}%
      \expandafter\def\csname LT6\endcsname{\color{black}}%
      \expandafter\def\csname LT7\endcsname{\color{black}}%
      \expandafter\def\csname LT8\endcsname{\color{black}}%
    \fi
  \fi
  \setlength{\unitlength}{0.0500bp}%
  \begin{picture}(6518.00,2436.00)%
    \gplgaddtomacro\gplbacktext{%
      \csname LTb\endcsname%
      \put(528,440){\makebox(0,0)[r]{\strut{} 0.5}}%
      \put(528,765){\makebox(0,0)[r]{\strut{} 1}}%
      \put(528,1090){\makebox(0,0)[r]{\strut{} 1.5}}%
      \put(528,1416){\makebox(0,0)[r]{\strut{} 2}}%
      \put(528,1741){\makebox(0,0)[r]{\strut{} 2.5}}%
      \put(528,2066){\makebox(0,0)[r]{\strut{} 3}}%
      \put(528,2391){\makebox(0,0)[r]{\strut{} 3.5}}%
      \put(934,220){\makebox(0,0){\strut{} 2180}}%
      \put(1482,220){\makebox(0,0){\strut{} 2200}}%
      \put(2030,220){\makebox(0,0){\strut{} 2220}}%
      \put(2578,220){\makebox(0,0){\strut{} 2240}}%
      \put(3126,220){\makebox(0,0){\strut{} 2260}}%
      \put(88,1415){\rotatebox{-270}{\makebox(0,0){\strut{}$I/I_{Si}$}}}%
      \put(2633,2098){\makebox(0,0)[l]{\strut{}(a)}}%
      \put(3259,0){\makebox(0,0){\strut{}Pressure (mbar)}}%
      \put(1345,2066){\makebox(0,0){\strut{}5.15 K}}%
      \put(1756,1741){\makebox(0,0){\strut{}5.16 K}}%
      \put(2304,1416){\makebox(0,0){\strut{}5.17 K}}%
      \put(2770,1220){\makebox(0,0){\strut{}5.18 K}}%
    }%
    \gplgaddtomacro\gplfronttext{%
    }%
    \gplgaddtomacro\gplbacktext{%
      \csname LTb\endcsname%
      \put(3787,440){\makebox(0,0)[r]{\strut{} 0.9}}%
      \put(3787,684){\makebox(0,0)[r]{\strut{} 1}}%
      \put(3787,928){\makebox(0,0)[r]{\strut{} 1.1}}%
      \put(3787,1172){\makebox(0,0)[r]{\strut{} 1.2}}%
      \put(3787,1416){\makebox(0,0)[r]{\strut{} 1.3}}%
      \put(3787,1659){\makebox(0,0)[r]{\strut{} 1.4}}%
      \put(3787,1903){\makebox(0,0)[r]{\strut{} 1.5}}%
      \put(3787,2147){\makebox(0,0)[r]{\strut{} 1.6}}%
      \put(3787,2391){\makebox(0,0)[r]{\strut{} 1.7}}%
      \put(3919,220){\makebox(0,0){\strut{} 0}}%
      \put(4741,220){\makebox(0,0){\strut{} 1000}}%
      \put(5563,220){\makebox(0,0){\strut{} 2000}}%
      \put(6385,220){\makebox(0,0){\strut{} 3000}}%
      \put(4166,2098){\makebox(0,0)[l]{\strut{}(b)}}%
      \put(4741,928){\makebox(0,0){\strut{}5.198 K}}%
      \put(4741,1903){\makebox(0,0){\strut{}5.25 K}}%
    }%
    \gplgaddtomacro\gplfronttext{%
    }%
    \gplbacktext
    \put(0,0){\includegraphics{120710_diffusion.eps_tex}}%
    \gplfronttext
  \end{picture}%
\endgroup